\documentclass[lettersize,journal]{IEEEtran}
\usepackage{amsmath,amsfonts}
\usepackage{algorithmic}
\usepackage{algorithm}
\usepackage{array}
\usepackage[caption=false,font=normalsize,labelfont=sf,textfont=sf]{subfig}
\usepackage{textcomp}
\usepackage{stfloats}
\usepackage{url}
\usepackage{verbatim}
\usepackage{graphicx}
\usepackage{cite}
\usepackage{booktabs}
\usepackage{multirow}
\usepackage{tabularx}
\usepackage[dvipsnames,table]{xcolor}
\usepackage{marvosym}

\newcommand{\mypara}[1]{\vspace{2pt}\noindent{\textbf{#1}}}

\begin{document}

\title{System-Level Isolation for Mixed-Criticality RISC-V SoCs: A "World" Reality Check}

\author{Luís Cunha$^\dagger$, José Martins $^\S$, Manuel Rodríguez$^\dagger$, Tiago Gomes$^\dagger$, and Sandro Pinto\\ Uwe Moslehner*, Kai Dieffenbach*, Glenn Farrall*, Kajetan Nuernberger*, and Thomas Roecker*\\ $^\dagger$Centro ALGORITMI/LASI, Universidade do Minho, $^\S$ OSYX Technnologies, *Infineon AG\vspace{-0.5cm}
\thanks{This work has been supported by FCT – Fundação para a Ciência e Tecnologia within the R\&D Unit Project Scope UID/00319/2025 - Centro ALGORITMI (ALGORITMI/UM) https://doi.org/10.54499/UID/00319/2025 and the Grant SFRH/BD/01007/2023.}%
}




\maketitle

\begin{abstract}

As RISC-V adoption accelerates, domains such as automotive, the Internet of Things (IoT), and industrial control are attracting growing attention. These domains are subject to stringent Size, Weight, Power, and Cost (SWaP-C) constraints, which have driven a shift toward heterogeneous Systems-on-Chip (SoCs) integrating general-purpose CPUs, tightly coupled accelerators, and diverse I/O devices with different integrity levels. While such integration improves cost efficiency and performance, it introduces a fundamental safety and security challenge: enforcing system-level isolation in mixed-criticality environments.
Although RISC-V International has proposed several hardware isolation primitives, including RISC-V Worlds, IOPMP, and SmMTT, their interoperability, scalability, and suitability for real-time systems remain insufficiently understood. In this paper, we present a comparative analysis of these primitives from the perspective of practical heterogeneous SoC designs. We implement an IOPMP, a World-based checker, and a modified RISC-V World checker that addresses key limitations of the baseline specification, and evaluate their trade-offs in terms of security guarantees and power–performance–area (PPA). Our results show that the World-based checker introduces a fixed, configuration-independent access latency, achieving lower worst-case delay than the evaluated alternatives while scaling predictably with system size. At the macro level, we estimate that the proposed modifications reduce SoC area by up to approximately 5\% compared to a baseline design. All artifacts will be released as open source, and we expect these findings to directly contribute to the evolution and ratification of RISC-V specifications, as well as to the design of future RISC-V SoCs.
\end{abstract}


\section{Introduction}
\IEEEPARstart{A}{s} RISC-V adoption accelerates, its scope is extending beyond general-purpose computing to encompass domains such as automotive, the Internet of Things (IoT), and industrial control. In the automotive sector in particular, RISC-V is poised to play a pivotal role across a broad range of devices, from edge Electronic Control Units (ECUs) to cross-domain and zonal controllers, Advanced Driver Assistance Systems (ADAS), and central compute platforms. These increasingly complex systems stand to benefit from the scalability and openness of the RISC-V ISA, which naturally aligns with the ongoing evolution of Electrical/Electronic (E/E) architectures toward centralized, software-defined designs~\cite{Staron2021, Mauser2024}.

\vspace{1mm}

Similar requirements are now emerging across industrial control and IoT markets~\cite{Cerrolaza2020}, where long product lifecycles, strong cost constraints, and stringent safety requirements reinforce the need for unified, industry-wide solutions. Meeting these requirements extends beyond the processor ISA itself and instead calls for a holistic compute platform that integrates a broad set of adjacent intellectual property (IP) blocks. This includes not only interrupt controllers, timers, and quality-of-service (QoS) mechanisms, but also robust isolation primitives capable of enforcing safety, security, and functional integrity across heterogeneous workloads and multiple levels of criticality. Ultimately, the ability to build and validate such platforms in a reliable and scalable manner will determine the viability of RISC-V in these domains.

\vspace{1mm}

Within this context, we identify two primary challenges. First, systems must provide robust isolation across mixed-criticality workloads, ensuring that faults or vulnerabilities in lower-integrity components (e.g., Quality Management (QM) workloads under ISO 26262) cannot compromise higher-integrity functions (e.g., ASIL-C/D workloads)~\cite{Palin2011, Cerrolaza2020,Gracioli2019,Ramsauer2017}. Second, systems must guarantee predictable and enforceable real-time quality-of-service (QoS), such that time-critical tasks maintain bounded latency even under resource contention~\cite{shedding_light,Zuepke2023,Costa2025,Lugo2022}. These challenges are particularly acute in cost-sensitive, high-volume platforms, where heterogeneous SoCs are employed to balance power efficiency and performance~\cite{Burns2018,Cinque2022,Mancuso2013}.


\vspace{1mm}

A key source of complexity in these platforms stems from the diversity of components and their interactions. In addition to CPUs and accelerators, modern SoCs integrate DMA engines, peripheral controllers, and GPUs, each with distinct privilege models, timing constraints, and protection requirements. In such environments, relying solely on initiator-side\footnote{We adopt the RISC-V nomenclature, referring to these entities as \emph{initiators} and \emph{targets}, replacing the legacy master and slave terminology.} protections managed by trusted firmware is increasingly inadequate: any firmware-level fault can undermine the entire safety model~\cite{rezone}. Moreover, initiator-side mechanisms are ill-suited to enforce system-wide policies across concurrently operating components interacting over shared interconnects~\cite{RodriguesTalk2023, Malka2015, Bonfils2024, siopmp}. For example, DMA engines assigned to user workloads can bypass initiator-side controls altogether, compromising SoC-wide integrity~\cite{Malka2015}. Consequently, ensuring system-wide protection requires complementary enforcement beyond the initiator, such as within the interconnect or at the target side.

\vspace{1mm}
Target-side protection in a computing platform typically comprises two core components: (i) initiator identification and (ii) target-side access checking. While the checking logic itself is comparatively straightforward, system-level initiator identification has historically relied on proprietary mechanisms, limiting both security guarantees and interoperability. For example, Arm defines multiple approaches for initiator identification, applies different schemes across its architectural profiles, and ultimately leaves their realization to product-specific implementations~\cite{Pinto2019}. Consequently, system-level isolation varies not only between application processors (Cortex-A) and microcontrollers (Cortex-M), but also within a single architectural profile. On Cortex-M–based platforms, for instance, isolation mechanisms may range from SAU/IDAU-based schemes to vendor-specific peripheral firewalls and interconnect checkers, each offering distinct semantics, granularity, and enforcement models.

Within the RISC-V ecosystem, several proposals have emerged to address these limitations, including RISC-V Worlds~\cite{worldguard}, Input/Output Physical Memory Protection (IOPMP)~\cite{iopmp}, and the more recent SmMTT~\cite{smmtt}. These efforts signal clear momentum toward standardized, system-level isolation mechanisms. However, while targeting partially distinct use cases, the coexistence of multiple alternative approaches introduces a degree of fragmentation that may be avoidable. Moreover, the interoperability among these mechanisms and their suitability for real-time systems remain largely unexplored, leaving open questions regarding their practical applicability and adoption.


In summary, current system-level isolation mechanisms are fragmented in both structure and enforcement. Existing designs differ significantly in how initiators are identified, how access-control policies are expressed, and where enforcement is applied within the system. These divergences arise not only across architectural profiles, but also within platforms based on the same profile. Consequently, isolation mechanisms pursuing similar objectives exhibit different latency, scalability, and security characteristics. While recent RISC-V proposals aim to standardize system-level protection, their practical trade-offs and deployment implications remain insufficiently understood.

In light of the fragmentation and open questions surrounding existing RISC-V system-level isolation mechanisms, this paper makes the following contributions:

\begin{itemize}
    \item a comparative analysis of RISC-V system-level isolation mechanisms, including RISC-V Worlds, SmMTT, and IOPMP, highlighting differences in initiator identification, isolation granularity, and permission management (Section~III);

    \item architectural extensions to the Worlds Checker that improve support for sparse memory layouts and scalability under large identifier counts, while preserving deterministic access behavior (Section~IV);
    
    \item a hardware implementation of three target-side protection primitives: IOPMP, the standard Worlds Checker, and a modified Worlds Checker (Section~V);
    
    \item a micro and macro evaluation of the implemented primitives integrated into a CVA6-based SoC, focusing on access latency, scalability, and hardware cost across checker configurations (Section~VI).
\end{itemize}

\section{Background \& Motivation}

\begin{figure}[t!]
    \centering
    \includegraphics[width=\linewidth]{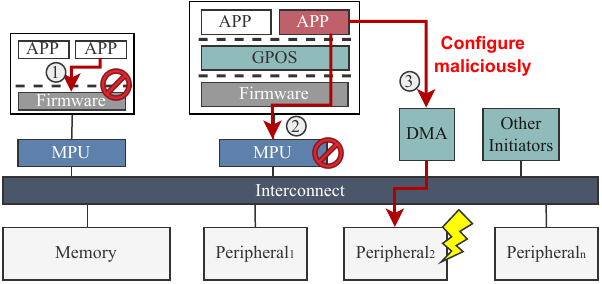}
    \caption{SoC overview illustrating the limitations of initiator-side protection.
(1) Trap-and-emulate at the CPU, (2) initiator-side access checks, and (3) a DMA-capable peripheral configured by an application to bypass protections.}
    \label{fig:soc_attack}
\end{figure}

Historically, initiator-side protection mechanisms have been deployed across a wide range of computing platforms, from high-performance application processor units (APUs) to resource-constrained microcontrollers (MCUs). These mechanisms span from full-featured memory management units (MMUs), which partition memory into page-granular regions across multiple execution contexts, to simpler memory protection units (MPUs) that enforce read, write, and execute permissions directly in the physical address space. However, such protections have traditionally been confined to CPU cores, leaving other system initiators without comparable enforcement. Modern SoCs integrate a diverse set of components, including GPUs, DMA engines, and peripheral controllers, each with distinct privilege models, timing constraints, and protection requirements. While techniques such as trap-and-emulate or paravirtualization can mitigate certain risks for user-level applications~\cite{Dall2014}, relying solely on initiator-side protections configured and managed by trusted firmware is increasingly inadequate in low-latency or safety-critical contexts~\cite{shedding_light, rezone}. In the absence of emulation, user-level software can further bypass these mechanisms by configuring devices (e.g., DMA engines) to access restricted memory on its behalf~\cite{Malka2015}, as depicted in Figure \ref{fig:soc_attack}. Together, these limitations underscore the need to extend protection mechanisms beyond the initiator and into other points of the system architecture.

Target-level protection typically relies on two main components: (i) initiator identification and (ii) target-side access checking. While target-side checking is fairly straightforward, initiator identification remains the least standardized element. 
Industry vendors have historically centered around Arm's TrustZone (TZ)~\cite{Pinto2019}. TZ extends the memory system by introducing an additional control signal, the non-secure (NS) bit, which propagates alongside memory and peripheral transactions to indicate their security state. This mechanism partitions the physical address space into secure and non-secure regions, forming the basis for access-control enforcement. However, while the architecture defines the notion of security states, it leaves their concrete enforcement largely implementation-defined.
As a result, vendors implement initiator identification and target-side enforcement in markedly different ways, directly limiting interoperability and portability across systems. 

Table~\ref{tab:platform_protections} summarizes representative implementations from major vendors, illustrating how different platforms instantiate these primitives, the number of supported security domains, and their corresponding device classes. In the following, we provide an overview of system-level isolation mechanisms available in mainstream APUs and MCUs, highlighting the significant heterogeneity and lack of interoperability that characterize current designs.

\begin{table*}[t!] 
    \caption{Comparison of vendor-specific system-level protection mechanisms.}
	\label{tab:platform_protections}
	\centering
	\scalebox{1}{%
	\begin{tabular}{
	    >{\centering\arraybackslash}p{1.5cm} 
	    >{\centering\arraybackslash}p{2cm} 
	    >{\centering\arraybackslash}p{3.5cm} 
	    >{\centering\arraybackslash}p{3cm} 
	    >{\centering\arraybackslash}p{3cm} 
        >{\centering\arraybackslash}p{2cm}} 
	
		\toprule
        \parbox{1.5cm}{\centering \small \textbf{Class}}
        & \parbox{2cm}{\centering \small \textbf{Vendor}}  
        & \parbox{3.5cm}{\centering \small \textbf{Platform}}
        & \parbox{3cm}{\centering \small \textbf{Initiator Identification}}
        & \parbox{3cm}{\centering \small \textbf{Target Checking}}
        & \parbox{2cm}{\centering \small \textbf{Number of Domains}}\\
        
        \midrule        
        \multirow{10}{*}{APU} 
        & \multirow{2}{*}{Renesas} 
        & RZ-T2H / RZ-N2H~\cite{rzt2h_rzn2h_um} RZ-Five~\cite{rzv_um}  
        & \footnotesize \multirow{2}{*}{TZ-Aware} 
        & \footnotesize \multirow{2}{*}{Multiple TZC-400} 
        & \multirow{2}{*}{2} \\
        \cmidrule(l){2-6}
        
        & \multirow{4}{*}{STMicroelectronics} 
        & \multirow{2}{*}{STM32MP1~\cite{stm32mp157_rm}} 
        & \footnotesize TZ-Aware + ETZC + Hardwired NSAID 
        & \footnotesize \multirow{2}{*}{ETZC + TZC-400} 
        & \multirow{2}{*}{2} \\
        \cmidrule(l){3-6}
        
        &  
        & STM32MP2~\cite{stm32mp23_rm} 
        & \footnotesize RIF 
        & \footnotesize RIF 
        & Up to 8 \\
        \cmidrule(l){2-6}
        
        & \multirow{3}{*}{NXP}  
        & i.MX 8QM~\cite{imx8qm_rm}
        & \footnotesize XRDC2 
        & \footnotesize XRDC2 
        & Up to 8 \\
        \cmidrule(l){3-6}

        &  
        & i.MX 9x~\cite{imx91_rm}
        & \footnotesize TRDC 
        & \footnotesize TRDC 
        & Up to 16 \\

        \midrule
        \multirow{17}{*}{MCU} 
        & \multirow{5}{*}{Renesas}  
        & Synergy~\cite{s7g2_um}  
        & \footnotesize Grouped Initiators 
        & \footnotesize Bus MPU 
        & n/a \\
        \cmidrule(l){3-6}
        
        &  
        & RA6M4~\cite{ra6m4_um} / RA6M5~\cite{ra6m5_um} / RA8M1~\cite{ra8m1_um}   
        & \footnotesize \multirow{2}{*}{MSAU + TZ-Aware} 
        & \footnotesize \multirow{2}{*}{TZ-Aware + Bus MPU}
        & \multirow{2}{*}{2} \\
        \cmidrule(l){3-6}
        
        &  
        & RH850/U2A~\cite{rh850_um}   
        & \footnotesize System-Protection ID 
        & \footnotesize Target-Side Filtering
        & Up to 32 \\
        \cmidrule(l){2-6}
        
        & STMicroelectronics 
        & \footnotesize STM32L/U/H5~\cite{stm32l5_rm,stm32u5_rm,stm32h5_rm} 
        & \footnotesize GTZC + TZ-Aware 
        & \footnotesize GTZC + TZ-Aware 
        & 2 \\
        \cmidrule(l){2-6}
        
        & Nordic 
        & nRF9160~\cite{nrf9160_ug} / nRF5340 
        & \footnotesize SPU 
        & \footnotesize SPU 
        & 2 \\
        \cmidrule(l){2-6}
        
        & Microchip 
        & SAM L1x~\cite{sam_l10_l11_ds}
        & \footnotesize IDAU 
        & \footnotesize PAC + TZ-Aware 
        & 2 \\
        \cmidrule(l){2-6}

        & \multirow{5}{*}{NXP}
        & LPC55Sxx~\cite{lpc55_um} / LPC552x 
        & \footnotesize MSW 
        & \footnotesize PPC + MPC 
        & 2 \\
        \cmidrule(l){3-6}

        &  
        & i.MX RT10xx~\cite{imxrt1010_rm}
        & \footnotesize CSU 
        & \footnotesize AIPSTZ 
        & 2 \\
        \cmidrule(l){3-6}

        &  
        & i.MX RT11xx~\cite{imxrt1170_rm}
        & \footnotesize RDC \& XRDC2 
        & \footnotesize RDC \& XRDC2
        & Up to 16 \\

        \cmidrule(l){2-6}
        & \multirow{4}{*}{Infineon}  
        & TVII
        & \footnotesize Protection Context 
        & \footnotesize Bus MPU
        & Up to 16 \\
        \cmidrule(l){3-6}
        
        &   
        & TC3
        & \footnotesize Master-TAGs 
        & \footnotesize Target-Side Filtering 
        & Up to 64 \\
        \cmidrule(l){3-6}

        &  
        & TC4
        & \footnotesize Master-TAGs + VM-IDs
        & \footnotesize Target-Side Filtering 
        & Up to 80 \\
        
	\bottomrule
	\end{tabular}
	}
\end{table*}




\mypara{Application Processor Unit.}
At the system level, APUs extend protection beyond CPUs by propagating initiator identity across the interconnect. In virtualized environments, Arm specifies the System Memory Management Unit (SMMU)~\cite{arm_smmu_v3}, which extends address translation and access control to non-CPU initiators. Each transaction carries a Stream ID (and Substream ID in SMMUv3) used to select a context descriptor defining the applicable translation tables and permissions. For non-virtualized systems, the TrustZone Address Space Controller (TZASC)~\cite{arm_tzc400} enforces access control using the Non-secure Access Identifier (NSAID) and associated security attributes (only the TZC-400 variant exposes the NSAID). Despite targeting different profiles, both mechanisms share a fundamental limitation: identifier assignment, semantics, and enforcement remain largely implementation-defined, limiting portability and interoperability.

The Arm standards do not specify how Stream IDs or NSAIDs are allocated or propagated, leaving each SoC vendor to define its own mapping. In practice, most implementations hardwire identifiers to initiators or interconnect ports, precluding dynamic reassignment.

\begin{itemize}
    \item In AMD Zynq UltraScale+ MPSoCs, each initiator is assigned a fixed 15-bit Stream ID~\cite{Costa2025}, while in STMicroelectronics STM32MP1 devices, NSAIDs are statically defined on a per-initiator basis~\cite{stm32mp157_rm}. As a consequence, even initiators operating within the same execution context must be represented by distinct identifiers, resulting in redundant configuration and limited scalability.

    \item TZASC do not support privilege reduction within firmware~\cite{rezone}. Although they restrict memory access, code executing at S-EL1 retains full control over their configuration and can reprogram them at runtime, allowing a compromised trusted OS to override system-wide permissions. Moreover, TZASC-based designs cannot distinguish between privilege levels, precluding isolation among software components within the secure world. While the SMMU offers finer-grained enforcement, its deployment in practice remains limited to a subset of bus initiators~\cite{rezone}.

    \item Vendors such as NXP, with the Trusted Resource Domain Controller (TRDC)~\cite{trdc}, and STMicroelectronics, with the Resource Isolation Framework (RIF)~\cite{rif}, have extended the TrustZone model beyond its dual-world abstraction by enabling multiple isolated domains through transaction tagging and target-side checks. While this improves isolation granularity, it remains incomplete. For instance, in NXP S32G processors, Cortex-A53 cores within a cluster share identical identifiers~\cite{nxp_xrdc}, preventing per-core isolation and, consequently, comprehensive system-level protection.

\end{itemize}

\mypara{Microcontroller Unit.}
At the lower end of the spectrum, fragmentation is even more pronounced across computing platforms. Prior to TrustZone-M in Armv8-M, no standardized platform-level isolation mechanism existed, and protection in Arm-based systems without vendor extensions relied entirely on software. Privileged firmware configured DMA descriptors, while the MPU restricted unprivileged access to peripheral registers, preventing unauthorized DMA setup but not memory access once DMA was active.
To improve isolation, several vendors introduced peripheral firewalls that shifted enforcement into hardware, adapting TrustZone-inspired concepts to non-TZ microcontrollers. For instance, the i.MX RT10xx family~\cite{imxrt1010_rm} integrates a Central Security Unit (CSU) that assigns privilege attributes to initiators, which are then enforced by the AHB-to-IP bridge (AIPSTZ).

Despite the security improvements introduced with Armv8-M, their implementation remains largely vendor-specific. In TrustZone, initiators tag transactions as secure or non-secure, and targets enforce access permissions accordingly. While TZ-aware components can natively generate and interpret these tags, non-TZ-aware peripherals remain unprotected. To address this gap, vendors insert custom interconnect modules to tag and check transactions on behalf of such peripherals. For example, in the RA6M4~\cite{ra6m4_um}, Renesas introduces Master Security Attribution Units (MSAUs) to tag initiator transactions and Bus MPUs to enforce target-side checks. Although effective, these manufacturer-specific solutions further contribute to ecosystem fragmentation.


Some platforms avoid explicit initiator identifiers altogether. For instance, Renesas’s Synergy Platform groups multiple initiators into clusters and assigns permissions at the cluster level~\cite{s7g2_um}. While this approach avoids the complexity of identifier management, it comes at the cost of granularity: any permission granted to a cluster applies uniformly to all its members, even when only a subset requires access. As a result, flexibility is reduced and isolation is weakened, since lower-integrity initiators may inherit privileges intended for higher-integrity ones.

Following an approach similar to those used in APUs, Infineon and Renesas (RH850 platform) employ dedicated initiator identifiers combined with target-side filtering mechanisms to enforce access permissions. Each on-chip resource, with direct or indirect bus access, is assigned an identifier that identifies the initiator issuing a transaction and is used to apply the corresponding access-control rules.

\section{System-Level Protection in RISC-V}

Target-side security guarantees are also a well-recognized requirement within the RISC-V ecosystem. Ongoing efforts seek to standardize initiator identification, with multiple alternative approaches currently under development. While these initiatives signal clear momentum toward standardized system-level isolation, their coexistence introduces fragmentation that risks hindering broad RISC-V adoption, echoing many of the pitfalls observed in proprietary Arm-based systems. The most prominent proposals in this space are RISC-V Worlds, SmMTT, and IOPMP. 

\begin{figure}[t!]
    \centering
    \includegraphics[width=\linewidth]{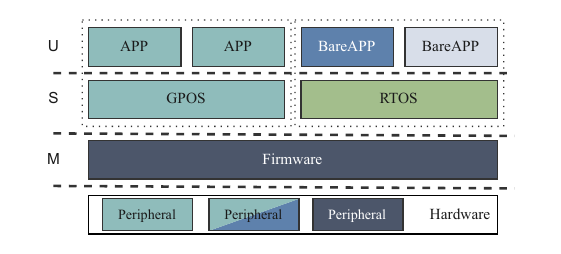}
    \caption{RISC-V Worlds system overview. Each color represents a different WID.}
    \label{fig:wg_system_level}
\end{figure}

\mypara{RISC-V Worlds.} RISC-V Worlds~\cite{worldguard} provides a system-wide isolation model based on the notion of worlds. Worlds represent execution contexts that encompass both initiators and targets and are each identified by a World Identifier (WID). RISC-V Worlds relies on the static allocation of agents and resources to worlds, typically configured by M-mode firmware or a trusted execution environment (TEE) at boot time. As illustrated in Figure~\ref{fig:wg_system_level}, WIDs can be assigned across the entire system stack, from user-mode applications to hardware components. To support this model, RISC-V Worlds introduces ISA extensions that allow cores to specify WIDs at different privilege levels, as well as non-ISA extensions in the form of the World Checker, which enforces target-side access control based on WIDs.

\mypara{SmMTT.} SmMTT~\cite{smmtt} specifies Supervisor Domain Access Protection, a RISC-V privileged architecture extension designed to isolate physical memory and device access. It introduces the Supervisor Domain Identifier (SDID), which tags hart-local resources and associates them with access-control metadata, thereby partitioning memory access across supervisor domains (i.e., software executing in S-mode). In addition, SmMTT defines a new security primitive, the Memory Protection Table (MPT), which can be used both at the hart level and within the I/O Bridge. Both the SDID and the MPT are managed by a trusted M-mode software monitor, the Root Domain Security Manager (RDSM), which constitutes a critical component of the trusted computing base (TCB). The RDSM operates analogously to a lightweight hypervisor or separation kernel, as it is responsible for (i) managing the MPT and (ii) performing context switches between supervisor domains. Moreover, the RDSM is the only hart-resident entity with unrestricted access to the entire physical address space.

\mypara{IOPMP.} Unlike RISC-V Worlds and SmMTT, the IOPMP specification~\cite{iopmp} focuses exclusively on target-side checking and leaves initiator-side tagging undefined. While it assumes the existence of propagated identifiers, namely Requester-Role Identifiers (RRIDs), to enforce access control on unprotected targets, the definition, assignment, and management of these identifiers are entirely implementation-specific. As a result, IOPMP inherits limitations similar to those observed in Arm-based systems, where proprietary tagging mechanisms constrain interoperability and scalability. Moreover, the absence of a standardized tagging model implies that IOPMP alone does not constitute a complete system-level isolation solution.

\subsection{Comparative Analyzes}
RISC-V system-level protection primitives differ substantially in how initiators are identified, how access checks are enforced, and how isolation domains are defined. Table~\ref{tab:rv_platform_protections} summarizes these distinctions for RISC-V Worlds, SmMTT, and IOPMP, and serves as the basis for the comparative analysis that follows.

\begin{table}[t!]
    \caption{Comparison of RISC-V proposed system-level protection mechanisms.}
    \label{tab:rv_platform_protections}
    \centering
    \footnotesize
    \setlength{\tabcolsep}{3pt}
    \renewcommand{\arraystretch}{1.1}
    \begin{tabularx}{\columnwidth}{
        >{\centering\arraybackslash}m{1.9cm}
        >{\centering\arraybackslash}X
        >{\centering\arraybackslash}X
        >{\centering\arraybackslash}X
        >{\centering\arraybackslash}m{1.6cm}}
        \toprule
        \multirow{2}{*}{\textbf{Specification}}
        & \multirow{2}{*}{\textbf{Granularity}}
        & \textbf{\scriptsize Initiator Identification}
        & \textbf{Target Checking}
        & \multirow{2}{*}{\textbf{Domains}} \\
        \midrule
        \multirow{2}{*}{RISC-V Worlds} & U-Mode Software & \multirow{2}{*}{World ID} & WC & \multirow{2}{*}{32 (128)} \\
        \midrule
        \multirow{2}{*}{SmMTT} & S-Mode Software & Supervisor Domain ID & \multirow{2}{*}{IOMPT} & \multirow{2}{*}{64} \\
        \midrule
        \multirow{2}{*}{IOPMP} & \multirow{2}{*}{Hardware} & Requester-Role ID & \multirow{2}{*}{IOPMP} & \multirow{2}{*}{n/a} \\
        \bottomrule
    \end{tabularx}
\end{table}

\mypara{Number of Domains.}
The number of supported domains directly impacts system scalability, as it bounds how many initiators can be independently controlled or partitioned within the architecture. Early versions of the RISC-V Worlds specify support for up to 32 WIDs, although recent proposals extend the maximum number of worlds to 128~\cite{security_hc_meeting}. SmMTT, in contrast, targets support for up to 64 domains. While IOPMP can support up to 65,535 RRIDs, this capacity does not directly translate into execution domains: RRIDs are used solely to identify the initiator issuing a transaction and are not intended to represent software execution contexts.

\mypara{Isolation Granularity.}
Granularity defines the smallest component that can be independently isolated and thus reflects the maximum isolation capability of an architecture. In RISC-V Worlds, isolation domains extend up to the user-level, enabling separation not only across OS domains but also among individual applications. In contrast, SmMTT defines isolation exclusively at the supervisor level; consequently, applications running within the same RTOS instance share a single domain, limiting achievable isolation. To mitigate this limitation, the specification anticipates the use of MMU/IOMMU mechanisms that further provide isolation between S- and U- modes. In IOPMP, by comparison, RRIDs are used solely to distinguish hardware initiators and are not associated with software execution contexts.

\mypara{Permission Management.}
In heterogeneous SoCs, high-integrity compute islands (e.g., hardware security modules) commonly handle security-critical functions such as access control, attestation, and asset management, while other harts execute lower-integrity workloads. In RISC-V Worlds, the WID for M-mode software can only be set by an external entity, such as a root of trust (RoT), or be hardwired. This entity may also configure all WCs at boot time and lock their configuration, thereby preserving the authority of the high-integrity island. Similarly, the IOPMP specification assumes checker configuration via a secure monitor, typically implemented as M-mode firmware or another high-integrity entity. In SmMTT, by contrast, SDID assignment and management of the MPT are handled by the RDSM executing on the hart. The MPT may be instantiated either globally, shared across all harts, or locally, with per-hart instances. While this flexibility simplifies configuration, particularly when multiple harts execute the same domain, it also introduces a systemic risk: a single fault or compromise in the RDSM can undermine isolation guarantees across the entire system.

\subsection{RISC-V Target-Side Checking Primitives}

Identifiers are only effective when coupled with hardware checkers that enforce access-control decisions before transactions reach memory or peripheral boundaries. These checkers act as the final enforcement layer and, despite their conceptual simplicity, their placement and internal design have a direct impact on latency, determinism, and area overhead, i.e., making them particularly critical in real-time heterogeneous SoCs.

Checker placement depends on the platform microarchitecture and desired trade-offs. Target-side placement typically minimizes the number of required entries, as protection scopes are limited to the associated resource. Initiator-side placement, by contrast, can act as early filtering by reducing or eliminating the need for identifier propagation, at the cost of flexibility and granularity. Such designs are therefore best suited to static deployments, but can improve resilience to denial-of-service attacks by protecting shared resources, such as the interconnect, earlier in the transaction path. Although checker placement is an important consideration, it lies largely outside the scope of this work. Our analysis focuses instead on the checker’s internal organization, scalability, and timing behavior, which are largely independent of its integration point within the system.

Within this design space, three RISC-V specifications define target-side checking primitives: IOPMP, the Worlds Checker (WC), and the Input/Output Memory Protection Table Checker (IOMPT). IOPMP operates independently of standardized initiator tagging, whereas WC and IOMPT extend the initiator-side mechanisms of RISC-V Worlds and SmMTT with corresponding target-side enforcement.

\mypara{IOPMP.}
The IOPMP specification defines three core components for transaction validation: (i) the Entry Array, (ii) the Memory Domain Configuration (MDCFG) table, and (iii) the Source-to-Memory Domain (SRCMD) table. Similar to PMP, each Entry Array element specifies a physical address range along with associated access permissions, which are evaluated to determine whether an incoming transaction is allowed. Entries are grouped into Memory Domains (MDs), enabling regions that share a common functional or security purpose to be managed collectively; this grouping is defined by the MDCFG table, which associates each entry with a specific MD. The SRCMD table then maps each RRID to one or more MDs. An RRID uniquely identifies a device, or a group of devices, that share the same access permissions.

\mypara{World Checker.}
Unlike IOPMP, the WC relies on a single structure, i.e., the slot, to validate transactions. Similar to an IOPMP entry, each slot defines a physical memory region and its associated access permissions. However, rather than distributing permissions across multiple tables, the WC encodes per-World permissions within a bitmap. For a given WID n in slot m, read and write permissions are stored at slot(m).perm[2n] and slot(m).perm[2n+1], respectively. When an incoming transaction matches the address range of a slot, the checker consults this bitmap to determine whether the access is permitted.

\mypara{IOMPT.}
Unlike the other two approaches, the IOMPT~\cite{smmtt} is not implemented as a standalone checker along the transaction path, but is instead integrated into the I/O Bridge's functioning. It employs a page-table–like structure that encodes access permissions for fixed 4 KiB physical memory regions. Upon receiving a transaction, the mechanism traverses the table in memory using the physical address until a leaf entry is reached. Depending on the address size, the leaf contains tuples specifying read, write, and execute permissions for the pages within the corresponding range. To mitigate lookup overhead, the IOMPT may employ caching structures to accelerate permission resolution. Moreover, because page tables can be shared across all IOMPT instances and the hart’s MPT, this design simplifies configuration and synchronization in homogeneous platforms. However, the reliance on memory-based table walks introduces higher latency and non-determinism, which can limit the suitability of IOMPT-based designs for real-time systems.

\section{Refined World-Checker: Overview} \label{sec:motivation}


Modern real-time and mixed-criticality SoCs increasingly rely on a large number of isolated execution contexts to separate applications with different trust and criticality requirements~\cite{wheels}. RISC-V Worlds addresses this need by providing a uniform system-level mechanism for access control across initiators and targets. However, as the number of isolation domains grows, several aspects of the current WC design raise scalability and efficiency concerns.


Early versions of the specification limited the WC to 32 WIDs. Recent proposals significantly increase this limit to 128 worlds via the \texttt{Smwdelegh} extension~\cite{release_128wids}, improving flexibility but exposing structural limitations in the checker design. In particular, the bitvector-based encoding of permissions scales poorly in hardware: wider vectors increase area and timing costs, even in systems where only a small subset of WIDs is relevant per initiator (e.g., classic master–satellite architectures). As a result, much of the added encoding capacity provides limited practical benefit while still incurring non-trivial overhead.
Additional inefficiencies stem from the address-matching model. In systems with sparse or fragmented memory layouts, protecting non-contiguous regions is costly: TOR matching consumes two slots per region, while NAPOT matching may waste address space due to alignment constraints. These limitations complicate incremental deployment, handover scenarios, and fine-grained protection of heterogeneous memory maps.

To address these issues, we propose a set of architectural refinements to the WC that improve scalability, flexibility, and efficiency. Specifically, we introduce: (i) enhancements to the slot structure (Tables~\ref{tab:curr_slot_struct} and~\ref{tab:slot_struct}), (ii) extensions to the configuration registers (Table~\ref{tab:cfg_struct}), and (iii) a redesigned permission encoding (Table~\ref{tab:perm_struct}):


\begin{itemize}
    \item \textbf{Start–End (SE) address matching (64-bit):}  
    We introduce an explicit start–end matching mode that complements TOR, NA4, and NAPOT. This mode enables precise region definition, reduces address-space waste, and better supports incremental configuration and handover use cases. To support SE matching, a new 64-bit \textit{eaddr} register stores the end address, while the existing address field now represents the start address. The slot configuration register’s address-mode field is extended by one bit to encode the new mode.

    \item \textbf{General-read permission bit:}  
    A general-read bit enables efficient sharing of read-only regions across multiple WIDs without duplicating permission entries. This bit is added to the configuration register (bit~24, Table~\ref{tab:cfg_struct}).

    \item \textbf{Explicit WID–permission entries:}  
    We replace the monolithic permission bitvector with explicit WID–permission pairs. Each slot may contain multiple such entries, allowing independent specification of permissions for different WIDs and improving scalability as the number of supported worlds increases (currently, up to 8 \textit{perm} fields). Table~\ref{tab:slot_struct} illustrates a configuration with four permission entries.

    \item \textbf{Slot field reorganization:}  
    The slot layout is reorganized to simplify the addition of multiple permission entries and to improve extensibility.

    \item \textbf{Expanded slot size:}  
    The slot size is increased from 32 to 64 bytes to accommodate the richer permission structure without increasing checker complexity.
\end{itemize}

\begin{table}[b!]
	\caption{Current specification's slot structure.}
	\label{tab:curr_slot_struct}
	\centering
	\scalebox{0.90}{%
	\begin{tabular}{
	   >{\centering\arraybackslash}p{1cm} 
	   >{\centering\arraybackslash}p{1.8cm}
	   >{\centering\arraybackslash}p{2.0cm}
	   >{\centering\arraybackslash}p{3.2cm}} 
	
		\toprule
        \centering \small \textbf{Offset}  
        & \parbox{1.8cm}{\centering \small \textbf{Bytes}}
        & \parbox{2.0cm}{\centering \small \textbf{Name}}
        & \parbox{3.2cm}{\centering \small \textbf{Description}} \\
        
        \cmidrule(l){1-4}
        \footnotesize 0x00 &   \footnotesize 4  & \footnotesize addr[33:2] & \footnotesize Rule address \\
        
        \cmidrule(l){1-4}
        \multirow{2}{*}{\centering\footnotesize 0x04} &   \multirow{2}{*}{\centering\footnotesize  4}  & \multirow{2}{*}{\centering\footnotesize  addr[65:34]} & \footnotesize Rule address (RV64 systems only) \\

        \cmidrule(l){1-4}
        \multirow{2}{*}{\centering\footnotesize  0x8} &   \multirow{2}{*}{\centering\footnotesize  8}  & \footnotesize perm[nWorlds-1:0] & \footnotesize Permissions for up to 32 worlds \\

        \cmidrule(l){1-4}
        \footnotesize 0x10 &   \footnotesize 4  & \footnotesize cfg & \footnotesize Rule Configuration \\

        \cmidrule(l){1-4}
        \footnotesize 0x14 &   \footnotesize 12  & \footnotesize n/a & \footnotesize Reserved \\

	\bottomrule
	\end{tabular}
	}
\end{table}

\begin{table}[b!]
	\caption{Proposed slot structure. New fields highlighted in bold.}
	\label{tab:slot_struct}
	\centering
	\scalebox{0.90}{%
	\begin{tabular}{
	   >{\centering\arraybackslash}p{1cm} 
	   >{\centering\arraybackslash}p{1.8cm}
	   >{\centering\arraybackslash}p{2.0cm}
	   >{\centering\arraybackslash}p{3.2cm}} 
	
		\toprule
        \centering \small \textbf{Offset}  
        & \parbox{1.8cm}{\centering \small \textbf{Bytes}}
        & \parbox{2.0cm}{\centering \small \textbf{Name}}
        & \parbox{3.2cm}{\centering \small \textbf{Description}} \\
        
        \cmidrule(l){1-4}
        \footnotesize 0x00 &   \footnotesize 8  & \footnotesize addr & \footnotesize Address rule, start address \\
        
        \cmidrule(l){1-4}
        \footnotesize \textbf{0x08} &   \footnotesize \textbf{8}  & \footnotesize \textbf{eaddr} & \footnotesize \textbf{Address rule, end address} \\

        \cmidrule(l){1-4}
        \footnotesize 0x10 &   \footnotesize 4  & \footnotesize cfg & \footnotesize Rule configuration \\

        \cmidrule(l){1-4}
        \footnotesize 0x14 &   \footnotesize 12  & \footnotesize n/a & \footnotesize Reserved \\

        \cmidrule(l){1-4}
        \centering\footnotesize \textbf{0x20} &   \footnotesize \textbf{4}  & \footnotesize \textbf{perm0} & \footnotesize \textbf{Permission configuration} \\

        \cmidrule(l){1-4}
        \footnotesize \textbf{0x24} &   \footnotesize \textbf{4}  & \footnotesize \textbf{perm1} & \footnotesize \textbf{Permission configuration} \\

        \cmidrule(l){1-4}
        \footnotesize \textbf{0x28} &   \footnotesize \textbf{4}  & \footnotesize \textbf{perm2} & \footnotesize \textbf{Permission configuration} \\

        \cmidrule(l){1-4}
        \footnotesize \textbf{0x2C} &   \footnotesize \textbf{4}  & \footnotesize \textbf{perm3} & \footnotesize \textbf{Permission configuration} \\

        \cmidrule(l){1-4}
        \footnotesize 0x30 &   \footnotesize 16  & \footnotesize n/a & \footnotesize Reserved \\
        
	\bottomrule
	\end{tabular}
	}
\end{table}

\begin{table}[t!]
	\caption{Perm field structure.}
	\label{tab:perm_struct}
	\centering
	\scalebox{0.90}{%
	\begin{tabular}{
	   >{\centering\arraybackslash}p{1cm} 
	   >{\centering\arraybackslash}p{1.8cm}
	   >{\centering\arraybackslash}p{2.0cm}
	   >{\centering\arraybackslash}p{3.2cm}} 
	
		\toprule
        \centering \small \textbf{Bits}  
        & \parbox{1.8cm}{\centering \small \textbf{Default}}
        & \parbox{2.0cm}{\centering \small \textbf{Name}}
        & \parbox{3.2cm}{\centering \small \textbf{Description}} \\
        
        \cmidrule(l){1-4}
        \footnotesize \multirow{2}{*}{6:0} &   \footnotesize \multirow{2}{*}{0}  & \footnotesize \multirow{2}{*}{wid} & \footnotesize World ID for the defined access permission \\
        
        \cmidrule(l){1-4}
        \footnotesize 29:7 &   \footnotesize n/a  & \footnotesize n/a & \footnotesize Reserved \\

        \cmidrule(l){1-4}
        \footnotesize 30 &   \footnotesize 1  & \footnotesize w & \footnotesize Write permission for WID \\

        \cmidrule(l){1-4}
        \footnotesize 31 &   \footnotesize 1  & \footnotesize r & \footnotesize Read permission for WID \\

	\bottomrule
	\end{tabular}
	}
\end{table}

\begin{table}[t!]
	\caption{Slot configuration structure. Only changes depicted in table.}
	\label{tab:cfg_struct}
	\centering
	\scalebox{1}{%
	\begin{tabular}{
	   >{\centering\arraybackslash}p{1cm}
	   >{\centering\arraybackslash}p{2.0cm}
	   >{\centering\arraybackslash}p{4cm}} 
	
		\toprule
        \centering \small \textbf{Bits}
        & \parbox{2.0cm}{\centering \small \textbf{Name}}
        & \parbox{4cm}{\centering \small \textbf{Description}} \\
        
        \cmidrule(l){1-3}
        \footnotesize 2:0  & \footnotesize A & \footnotesize Address Range Specification \\
        
        \cmidrule(l){1-3}
        \footnotesize 24 & \footnotesize GR & \footnotesize General read permission \\

	\bottomrule
	\end{tabular}
	}
\end{table}

\section{Design and Implementation}

Although the mechanisms discussed in the previous sections pursue similar isolation objectives, their structural differences lead to distinct trade-offs that vary with the application context. To study these differences, we designed and implemented three target-side protection primitives: IOPMP, the standard Worlds Checker (S-WC), and a modified Worlds Checker (M-WC) incorporating the extensions proposed in Section~\ref{sec:motivation}.

Given our focus on real-time suitability, we excluded the IOMPT. IOMPT-based designs store access permissions in a memory-resident MPT indexed by physical addresses, with enforcement performed by an MPT checker integrated into the core or an I/O Bridge. Such table-based mechanisms are ill-suited for real-time systems, as table walks introduce input-dependent latency unless supported by large lookaside buffers~\cite{Koenig2025}, which complicate worst-case execution-time analysis~\cite{Kahlen2025} and can negatively impact predictability and area efficiency, as additional QoS mechanisms may be needed.

\subsection{IOPMP Table Formats}

To improve flexibility, the IOPMP specification~\cite{iopmp} defines three alternative formats for both the MDCFG and SRCMD tables, enabling implementations to trade off area, latency, and configuration complexity. The available formats and their characteristics are summarized in Tables~\ref{tab:mdcfg_fmt} and~\ref{tab:srcmd_fmt}.

The implemented IOPMP supports SRCMD format~0 and MDCFG format~1. MDCFG formats~0 and~2 were omitted, as the tables are assumed largely static at runtime and formats~1 and~2 offer equivalent checking behavior. SRCMD format~1 was excluded because it does not support sharing Memory Domains (MDs) across multiple RRIDs. While full sharing is unnecessary for our use cases, limited sharing is essential in resource-constrained systems; in this respect, SRCMD format~2 would also be suitable, as it implicitly associates all RRIDs with all MDs.

\begin{table}[t!]
	\caption{MDCFG Formats and their description.}
	\label{tab:mdcfg_fmt}
	\centering
	\scalebox{.98}{%
	\begin{tabular}{
	   >{\centering\arraybackslash}p{1cm} 
	   >{\centering\arraybackslash}p{7cm}} 
	
		\toprule
        \centering \small \textbf{Format}  
        & \parbox{7cm}{\centering \small \textbf{Description}} \\
        
        \cmidrule(l){1-2}
        \footnotesize \multirow{3}{*}{0} &  \footnotesize Each MD(m) has a dynamic number of entries, where \textit{MDCFG(m).t} is the exclusive upper bound and \textit{MDCFG(m-1).t} is the inclusive lower bound (for $m > 0$) \\
        
        \cmidrule(l){1-2}
        \footnotesize \multirow{2}{*}{1} & \footnotesize No MDCFG table. All MDs share the same fixed number $k$ of entries  \\

        \cmidrule(l){1-2}
        \footnotesize \multirow{2}{*}{2} & \footnotesize SRCMD table holds per-RRID bitmap permissions over each MD; RRIDs are implicitly associated with all MDs.  \\

	\bottomrule
	\end{tabular}
	}
\end{table}

\begin{table}[t!]
	\caption{SRCMD Formats and their description.}
	\label{tab:srcmd_fmt}
	\centering
	\scalebox{.98}{%
	\begin{tabular}{
	   >{\centering\arraybackslash}p{1cm} 
	   >{\centering\arraybackslash}p{7cm}} 
	
		\toprule
        \centering \small \textbf{Format}  
        & \parbox{7cm}{\centering \small \textbf{Description}} \\
        
        \cmidrule(l){1-2}
        \footnotesize \multirow{2}{*}{0} &  \footnotesize Each SRCMD(\textit{m}) has a bitmap, mapping which MDs belong to a specific RRID \textit{m}.\\
        
        \cmidrule(l){1-2}
        \footnotesize \multirow{2}{*}{1} & \footnotesize No SRCMD table. RRIDs directly map to the MDs in a one-to-one relation. \\

        \cmidrule(l){1-2}
        \footnotesize \multirow{2}{*}{2} & \footnotesize No MDCFG table. All MDs share the same number $k$ of entries, which can be reprogrammed.  \\

	\bottomrule
	\end{tabular}
	}
\end{table}

Notably, SRCMD format~2 can be viewed as an attempt to align the IOPMP model with the WC approach. In this configuration, the SRCMD table mimics the WC permission field by employing a bitmap-based encoding, with permissions assigned at the MD level rather than per entry. From this perspective, format~2 represents a partial standardization of concepts introduced by RISC-V Worlds, albeit through a different structural abstraction. However, since our evaluation already includes the S-WC, comparing it against the baseline IOPMP configuration yields a clearer assessment of their respective design trade-offs.

\subsection{Implementation considerations}

Given our focus on real-time systems, latency constitutes the primary design constraint. Typical workloads in such systems consist of word-sized transactions to memory-mapped peripherals, often issued in irregular or bursty patterns. Moreover, these accesses frequently target non-cached memory regions, making any additional latency introduced along the transaction path directly observable. As a result, checker implementations must minimize per-access overhead and avoid variable or input-dependent delays. In many safety-critical embedded applications, system performance is ultimately bounded by worst-case latency rather than aggregate throughput, further reinforcing this requirement.

\begin{figure}[t!]
    \centering
    \includegraphics[width=1\linewidth]{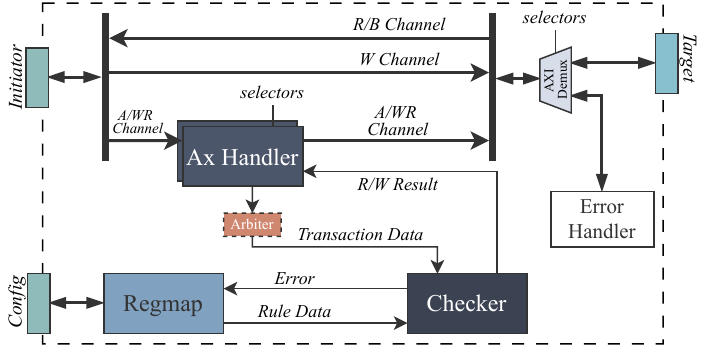}
    \caption{IOPMP and WC diagram overview.}
    \label{fig:checker_diagram}
\end{figure}

The implemented IOPMP and Worlds Checker variants share a common architecture, differing mainly in the checker module. As shown in Figure~\ref{fig:checker_diagram}, initiators connect via the Initiator AXI port and protected targets via the Target AXI port (referred to in IOPMP as the Requestor and Receiver ports, respectively). Dedicated Ax handlers for each channel (i.e. read and write) extract key transaction attributes, such as operation type, transfer length, base address, and initiator identifier (IID)\footnote{We use \emph{Initiator Identifier (IID)} to denote both World Identifiers (WIDs) and Requester-Role Identifiers (RRIDs); \emph{rule} refers generically to protection entries.}, and forward them to the checker for validation (through a round-robin arbiter). The result is then used to route the transaction, via PULP’s AXI demultiplexer, either to an error handler or to the target AXI channel.

\subsection{Checker}

The checkers across the three IPs share a common structure: transaction metadata from the Ax handlers is evaluated by a set of analyzers, each responsible for comparing the requested address against a rule. This design enables parallel rule evaluation, allowing a transaction to be validated in a single clock cycle when the number of analyzers matches the number of rules. While this minimizes latency, it incurs higher area and power overhead than sequential designs due to hardware replication across analyzers~\cite{Cunha2024}. Because rule structures differ across specifications, the checker module is implemented differently in each IP.

\mypara{IOPMP.}
The IOPMP checker adopts a pipelined design. In the first stage, it determines the set of rule indices to be evaluated; in the second stage, analyzers assess the validity of the transaction. If no matching rule is found for the transaction address within the current MD, and the IID is associated with multiple MDs, the checker stalls and re-invokes the second stage with the rule indices of the next eligible MD. This process repeats until a match is found or no further rules remain (invalid transaction).
This design introduces important implications for IOPMP analysis. Because the checker processes one MD per cycle, the number of analyzers needs only to match the number of rules within a single MD, rather than across all domains. While this approach increases worst-case latency, it avoids a substantial increase in logical depth that would otherwise result from the sequential evaluation of the multi-table structure. In practice, the latency impact can be mitigated by configuring the most frequently accessed rules in lower-indexed MDs.

\mypara{WG-Checker.} The WC is fully combinational, enabled by its flat rule structure and the independence of individual rules. This design allows address matching and WID-based permission checking to be performed in parallel. In contrast, the IOPMP requires accessing the SRCMD table to determine applicable rule indices, introducing inherent sequentiality. The absence of priority resolution in the WC further simplifies the logic and facilitates efficient parallelization. Our proposed modifications primarily target this module: rather than evaluating permissions through a bitmap, the WID is compared in parallel against all permission fields, and the corresponding access rights are selected upon a match.

\subsection{Rule Violation}

After the checker completes its evaluation, the result is forwarded to the Ax handlers, which determine the transaction’s final outcome. Permitted transactions are forwarded through the Target AXI port to the protected resource, while violations are routed to the Error Handler, which generates the appropriate response for the initiator. Two response types are supported: a bus error or poisoned data. Both specifications provide per-rule control bits to enable or suppress error signaling. Accordingly, the Error Handler either issues a bus error (e.g., an AXI4 \texttt{DECERR}) or discards write data and returns invalid read data, depending on the transaction type and the configuration of the matching rule.

\subsection{Rule Update}

In both specifications, the checkers are configured via memory-mapped registers. Rule address registers store encoded address values whose interpretation depends on the \textit{cfg.a} field, which is shared between IOPMP and the WC. To optimize timing within the checker, address decoding is performed at configuration time: the encoded value is decoded once and stored internally in start–end form. This simplifies the analyzer logic, which then only compares decoded bounds against transaction parameters.
However, this approach introduces configuration latency, temporarily decoupling the programmed permissions from the active system state. To mitigate this, the IOPMP specification defines a mechanism to stall transactions during reconfiguration, although this feature is not yet implemented in our IP.

\section{Evaluation}

All proposed IPs were integrated into a CVA6-based SoC and used to control memory accesses issued by an in-house DMA engine. For each checker, we evaluated multiple configuration scenarios to assess behavior under varying conditions. Validation was performed using both a verilated simulation environment and deployment on a low-cost FPGA platform (Genesys~2). In addition, we leveraged and expanded an existing bare-metal test framework \cite{Sa2022} to functionally validate each checker across the evaluated configurations.

\subsection{Latency} 
To quantify the latency overhead, we designed a worst-case scenario in which a DMA initiator issues single-beat requests, each transferring a single word. This configuration eliminates outstanding and out-of-order behavior, ensuring that the measured latency directly reflects the impact of the checkers under representative access patterns, such as peripheral accesses. We measure latency in clock cycles from request issuance to data availability at the initiator (for loads) or at the target (for stores). For IOPMP, best-case scenario (BCS) measurements were obtained using a single MD, corresponding to its optimal configuration, as discussed later.

\begin{figure}[t!]
    \centering
    \includegraphics[width=\linewidth]{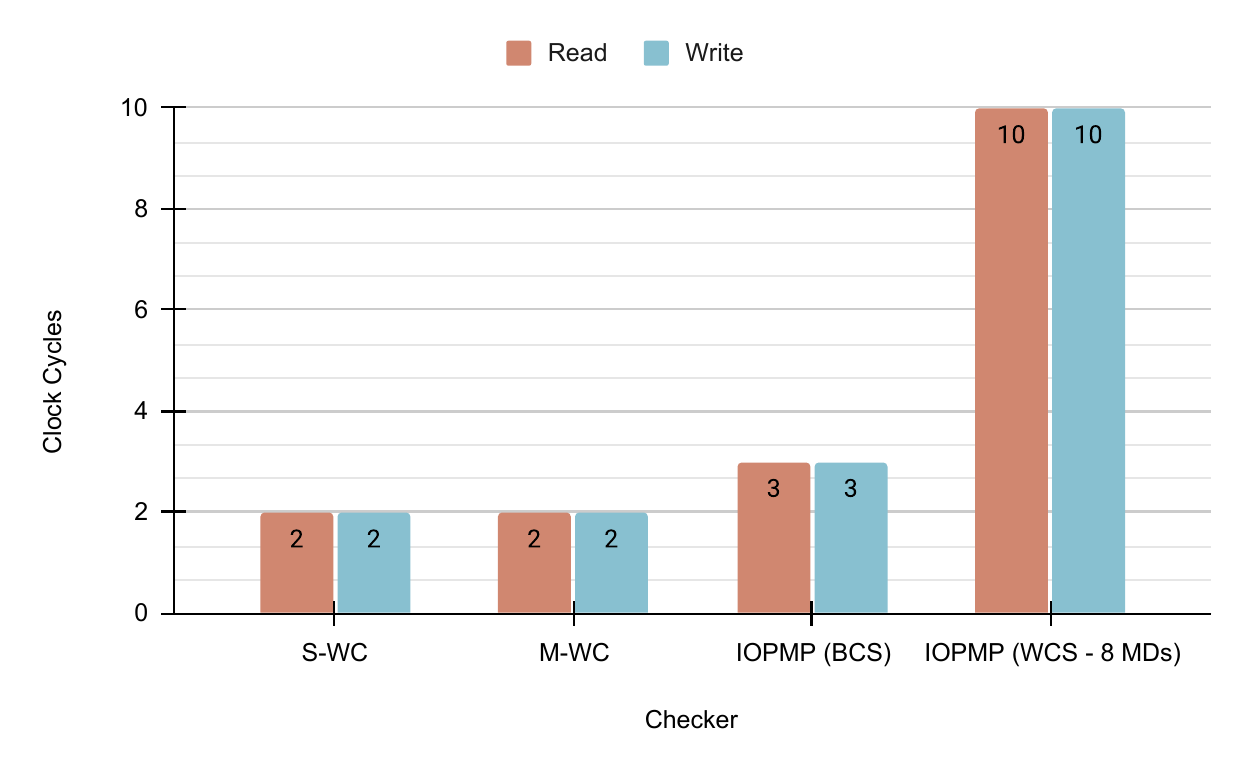}
    \caption{Introduced latency, in clock cycles, from the different IPs in read and write transactions.}
    \label{fig:latency_chart}
\end{figure}

Figure~\ref{fig:latency_chart} summarizes the measured checker latencies across configurations for both read and write transactions. Both the S-WC and the M-WC introduce identical overheads, adding two cycles per transaction. For the IOPMP, the best-case scenario incurs an additional three-cycle delay relative to the baseline. Notably, IOPMP latency depends on the position of the matching MD: if the IID is associated with \textit{n} MDs and the matching rule resides in the \textit{n}-th MD , the total latency increases by an additional $n – 1$ cycles. This behavior is illustrated in the final columns of the chart, which correspond to a match occurring in the eighth MD.

\begin{figure*}[t!]
    \centering
    \subfloat[LUT comparison between S-WC, M-WC, and IOPMP with different numbers of rules.]{
        \includegraphics[width=0.48\textwidth]{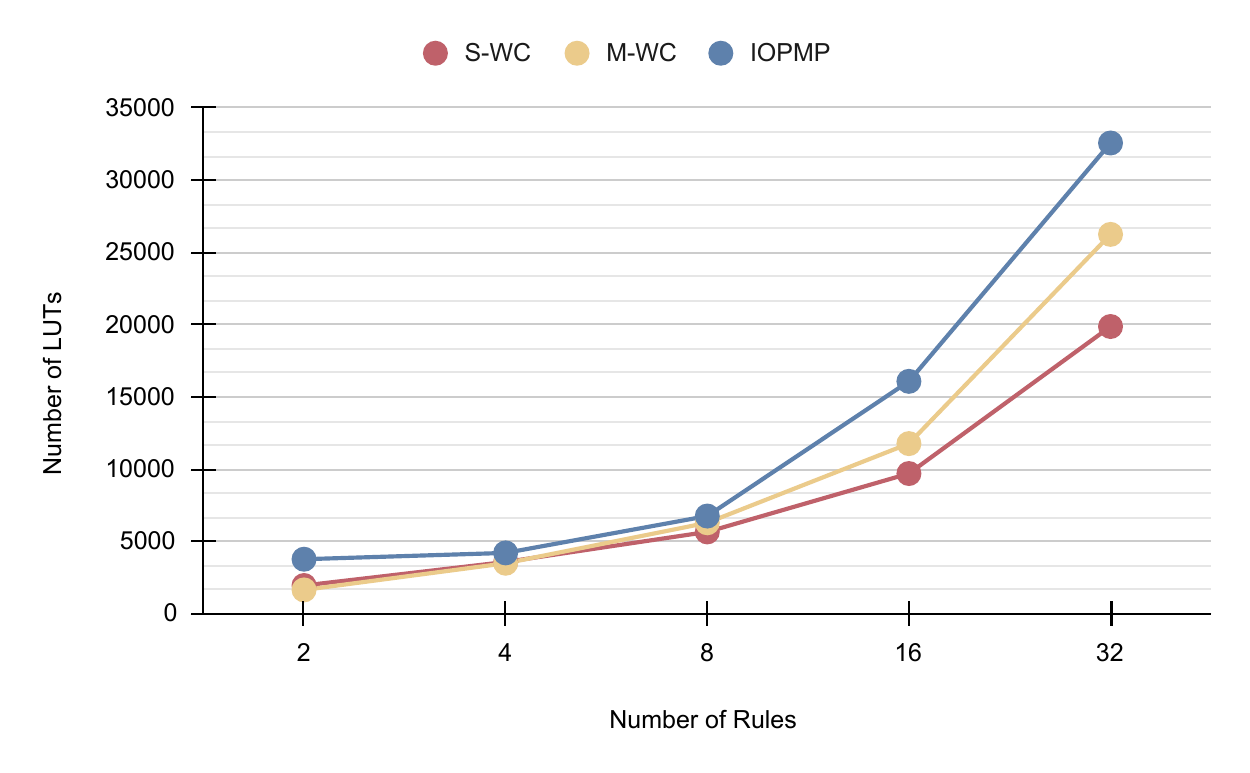}
        \label{fig:general_chart_lut}
    }
    \hfill
    \subfloat[FF comparison between S-WC, M-WC, and IOPMP with different numbers of rules.]{
        \includegraphics[width=0.48\textwidth]{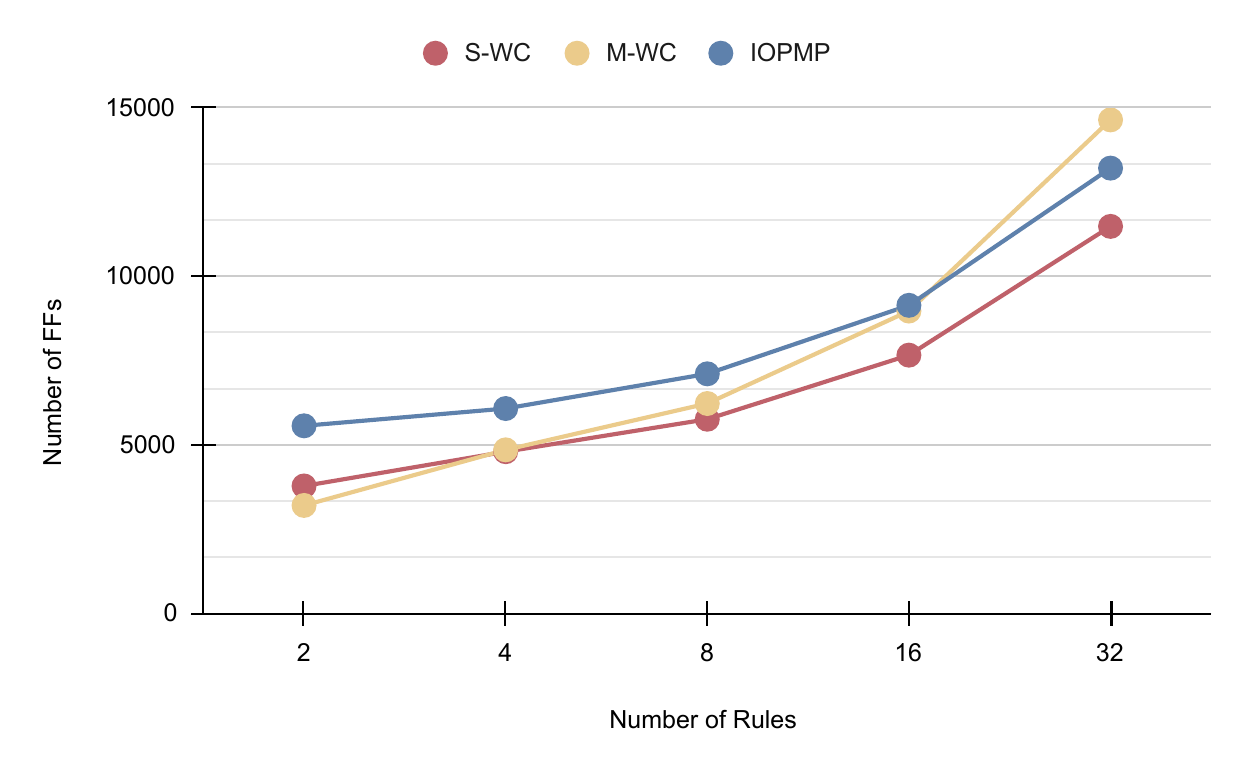}
        \label{fig:general_chart_ff}
    }

    \vspace{0.8em}

    \subfloat[LUT comparison between S-WC, M-WC, and IOPMP with different numbers of IIDs.]{
        \includegraphics[width=0.48\textwidth]{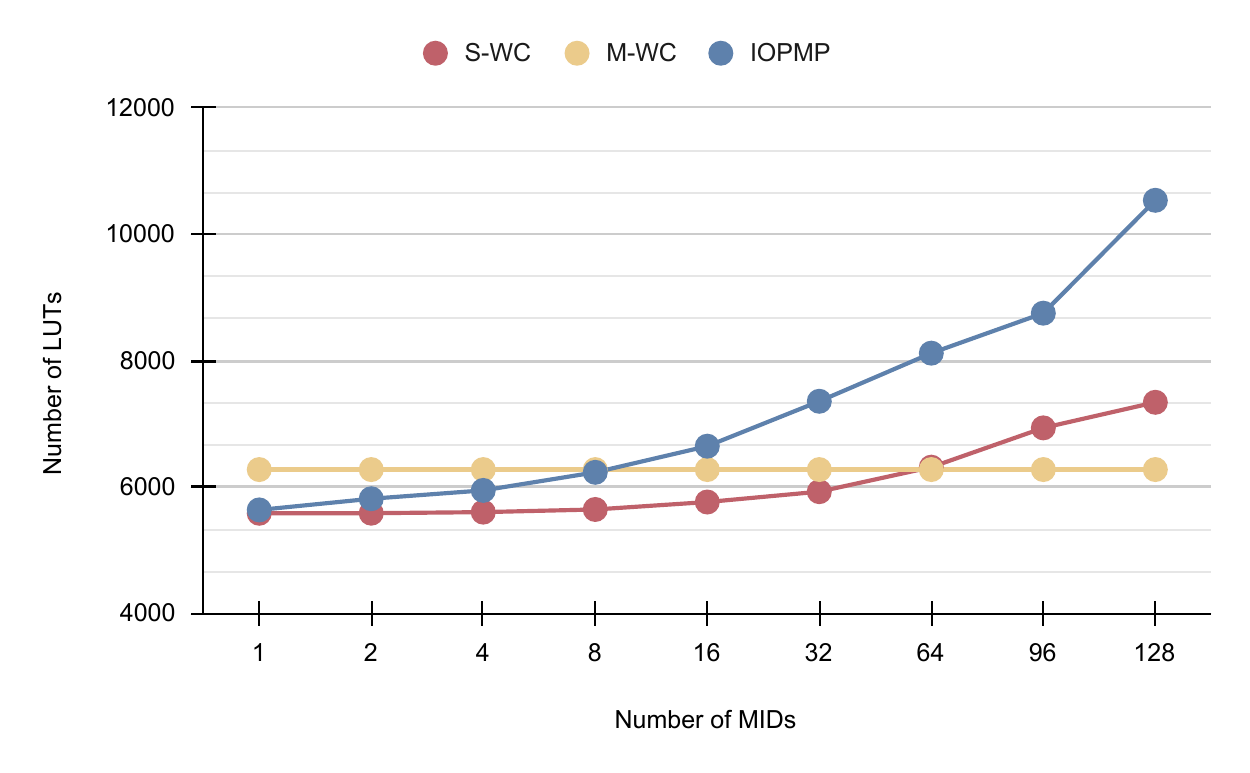}
        \label{fig:id_chart_lut}
    }
    \hfill
    \subfloat[FF comparison between S-WC, M-WC, and IOPMP with different numbers of IIDs.]{
        \includegraphics[width=0.48\textwidth]{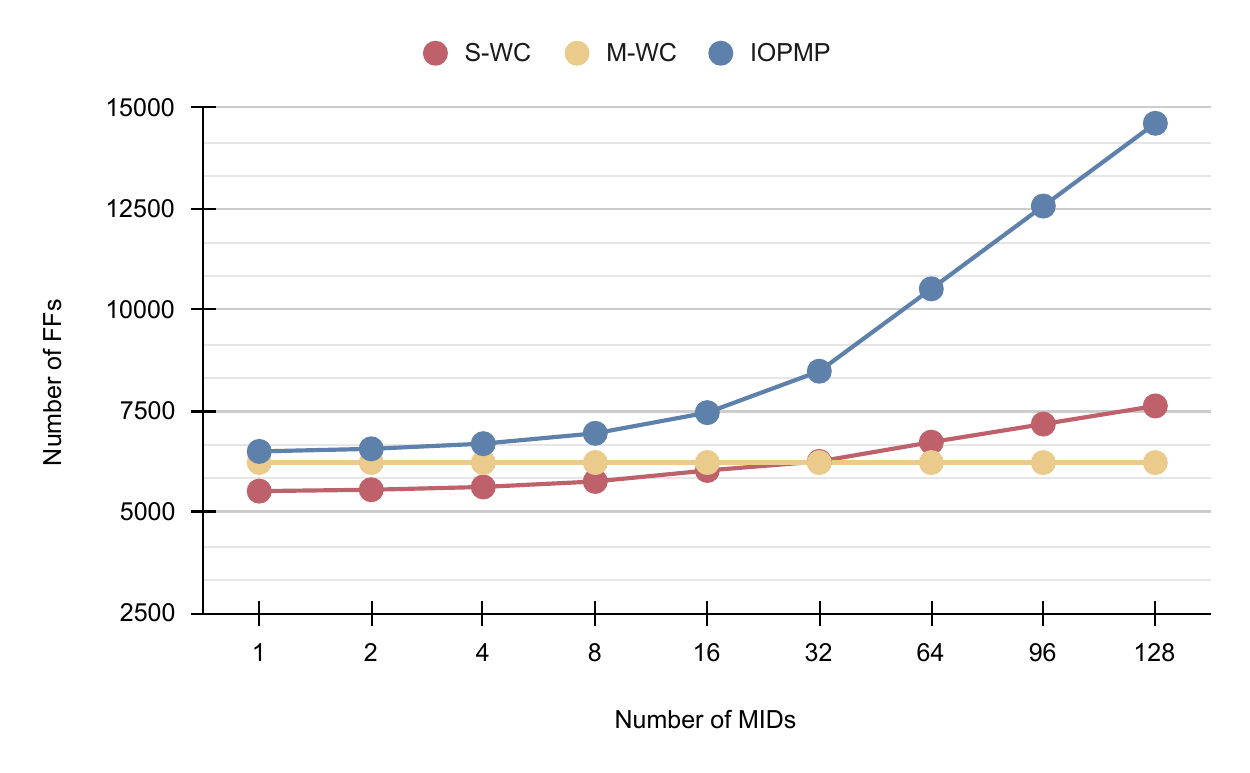}
        \label{fig:id_chart_ff}
    }

    \caption{Resource utilization comparison between S-WC, M-WC, and IOPMP across different rule and IID configurations. The top row shows scaling with the number of rules, while the bottom row highlights scaling with the number of IIDs.}
    \label{fig:wgc_resource_comparison}
\end{figure*}

\subsection{Hardware Resources} 
Resource utilization is highly dependent on both the synthesis tools and the target platform. Although the presented results do not target actual silicon, the evaluation methodology provides a consistent basis for relative comparison. To provide meaningful insight, we evaluated multiple configurations across different implementation scenarios. Our analysis compares overall resource consumption among the IPs under equivalent setups and examines how usage scales with the number of supported rules, IIDs, and sharing capabilities.

\mypara{Rule Scalability.} Figures~\ref{fig:general_chart_lut} and~\ref{fig:general_chart_ff} show the LUT and FF utilization of the evaluated IPs as the number of rules increases, with the number of IIDs fixed at eight. For small configurations (up to approximately eight rules), both Worlds Checker variants exhibit a clear area advantage over the IOPMP. With only two slots, the WC consumes nearly half the resources of a comparable IOPMP implementation. This behavior results from constraints imposed by the WC specification, which fixes the address ranges of the first and last rules to the lower and upper bounds of the protected address space, respectively. In addition, the first rule cannot be enabled (its only valid configuration is OFF), and the last rule is restricted to TOR mode. Consequently, these rules are effectively static, and the hardware normally required to support dynamic reconfiguration is unused, yielding a clear resource advantage in low-rule configurations.
As the number of rules increases, the area scaling of the two architectures becomes comparable. Overall, these results show that WC-based designs offer a clear resource advantage in minimal configurations, while their area consumption gradually converges with that of IOPMP as the number of rules increases.




\mypara{IID Scalability.} Figures~\ref{fig:id_chart_lut} and~\ref{fig:id_chart_ff} report the LUT and FF utilization, respectively, as the number of supported IIDs increases, with the number of rules fixed at eight. As the IID count grows, the scalability limitations of the bitmap-based permission encoding used in the S-WC become very clear. This effect would be even more pronounced with a larger rule set, since each additional IID introduces two extra bits per rule, excluding mapping and routing overhead. Although the S-WC maintains the lowest FF utilization up to 32 IIDs, its resource usage converges with that of the M-WC at this intersection. Although the specification currently limits support to 32 IIDs, the planned expansion to 128 worlds \cite{security_hc_meeting} motivated us to extend the S-WC beyond this limit. The IOPMP exhibits the poorest scalability in both LUT and FF usage. This behavior is largely driven by (i) the specification, which requires at least one additional 8-byte register per IID, and by (ii) the current IP design, which evaluates the SRCMD and MDCFG structures in an almost combinational fashion and would benefit from deeper pipelining. In contrast, the M-WC demonstrates relatively flat resource utilization as the number of IIDs increases. This behavior stems from its limited support for rule sharing, a trade-off that we analyze further in the following discussion.

\subsection{Permission Encoding and Scalability Trade-offs}

To assess the impact of the proposed WC extensions, we examine how each modification contributes to our two primary objectives: (i) improved support for sparse memory layouts and (ii) better scalability under large IID counts. With respect to sparse layouts, the introduced start–end address encoding directly addresses prior limitations by allowing arbitrary regions to be expressed without duplicating rules (previously required when using TOR) or wasting memory space for non–naturally aligned memory ranges.

Regarding scalability, the proposed changes increase the baseline hardware resource cost of the M-WC relative to the S-WC as the number of rules increases, as expected from the introduction of additional registers. Nevertheless, both designs exhibit similar scaling trends with respect to rule count. As shown in Figures~\ref{fig:id_chart_lut} and~\ref{fig:id_chart_ff}, a crossover point is observable when the number of IIDs increases while the number of rules remains fixed, beyond which the M-WC becomes increasingly more resource-efficient.

\begin{figure}
    \centering
    \includegraphics[width=\linewidth]{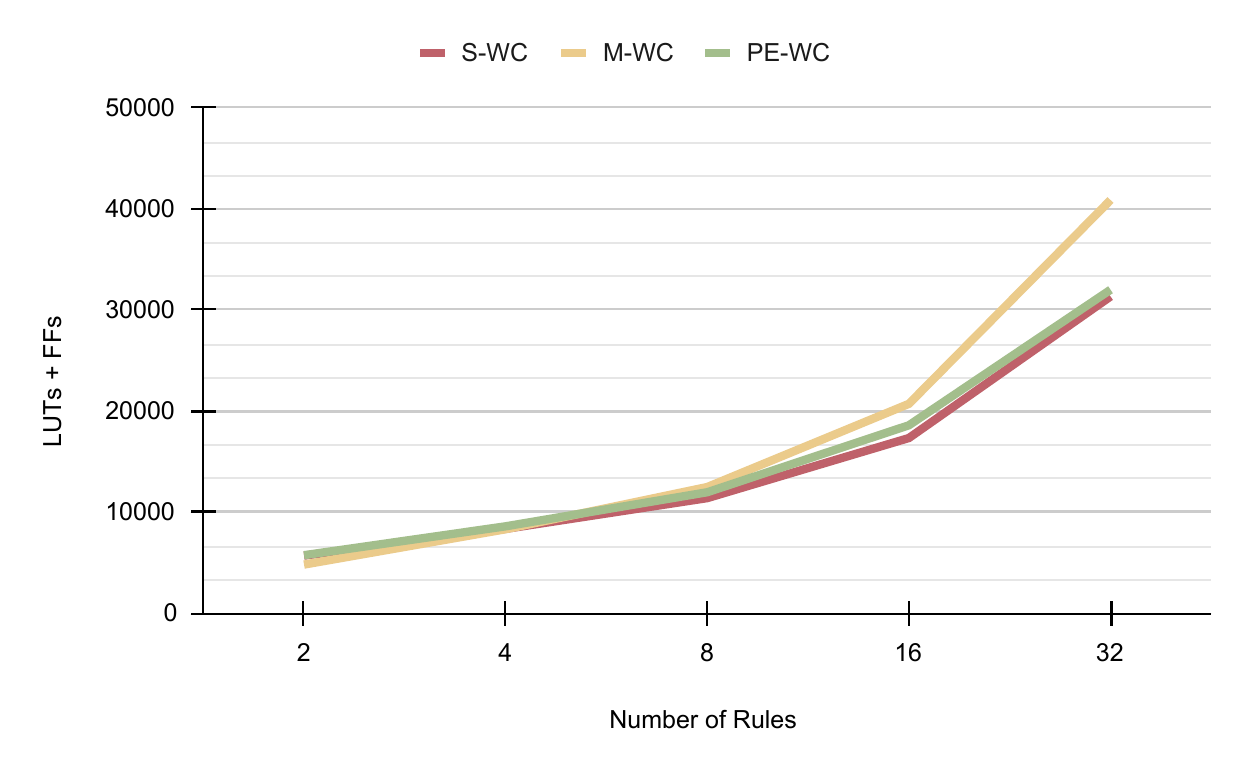}
    \caption{Total consumed resource comparison between the S-WC, M-WC, and PE-WC, as the number of rules increases (IIDs fixed at eight.)}
    \label{fig:study_slot_chart}
\end{figure}

\begin{figure}
    \centering
    \includegraphics[width=.965\linewidth]{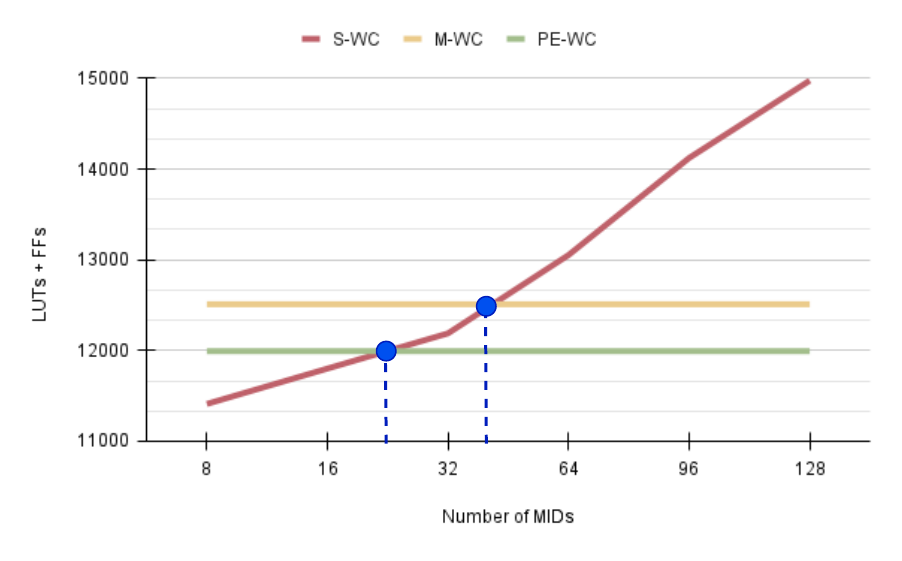}
    \caption{Total resource comparison between the S-WC, M-WC, and PE-WC, as the number of IIDs increases (rules fixed at 8). Highlighted in blue are the intersection points between the S-WC and the other implementations.}
    \label{fig:study_id_chart}
\end{figure}

To isolate the effect of the redesigned \textit{perm} field, we implemented an additional variant that preserves the S-WC structure while adopting the proposed permission encoding (PE-WC). Figures~\ref{fig:study_slot_chart} and~\ref{fig:study_id_chart} compare the total resource usage of all three implementations, enabling a more precise analysis of the shareability trade-off introduced by the new design.

As demonstrated in Figure~\ref{fig:study_slot_chart}, this configuration allows the PE-WC to retain virtually the same baseline resource usage as the S-WC, while achieving improved scalability as the number of IIDs increases (Figure~\ref{fig:study_id_chart}). With this reduced baseline, the crossover point at which the modified design becomes more resource-efficient, shifts from approximately 46 IIDs (in the S-WC versus M-WC comparison) to around 24 IIDs. This shift demonstrates that the redesigned \textit{perm} field is the primary contributor to the observed scalability improvements, enabling the modified WC designs to become more resource-efficient in systems with large IID counts while still preserving the compact footprint of the standard design for smaller configurations.

\subsection{SoC-level Impact Estimation}

To obtain a coarse-grained estimate of the system-level impact of the proposed optimizations, we complemented our micro-level evaluation with an SoC-level impact analysis. As a baseline, we used a CVA6-based design featuring a single hart and a minimal set of peripherals; for its memory layout, the smallest checker configuration deemed sufficient requires 16 rules. A checker configuration with 16 rules incurs an area overhead of approximately 15\%, which we use as a reference point for extrapolating resource consumption in larger systems. Although more complex designs typically require additional rules, resource usage is expected to scale proportionally. For consistency and comparative analysis, we therefore treat this 15\% overhead as the baseline for a 64-rule configuration. The resulting estimates are summarized in Table~\ref{tab:deployment_impact}.

As shown in Table~\ref{tab:deployment_impact}, the PE-WC scales with increasing IID counts, achieving area reductions of up to approximately 5\% in configurations with large numbers of rules and identifiers. In contrast, the M-WC incurs a higher baseline cost due to the start–end address encoding, causing this overhead to become more pronounced as the rule count increases and limiting its gains relative to the standard design. In more complex systems, where address spaces are typically fragmented, the start–end encoding can reduce (depends on deployment) the total number of required rules, for example by expressing non–naturally aligned regions with a single rule, thereby increasing the relative advantage of the M-WC over the S-WC. Overall, these results indicate that the proposed WC designs remain viable for deployment, with PE-WC offering the most favorable scaling behavior and M-WC becoming advantageous in more complex configurations with large IID counts and fragmented memory spaces.

\begin{table}[t!]
	\caption{Impact percentage of S-WC, PE-WC, and M-WC for different rule and ID configurations on an SoC.}
	\label{tab:deployment_impact}
	\centering
	\scalebox{0.90}{%
	\begin{tabular}{
	    >{\centering\arraybackslash}p{1cm} 
	    >{\centering\arraybackslash}p{1cm}
	      >{\centering\arraybackslash}p{1.8cm}
	      >{\centering\arraybackslash}p{1.8cm}
        >{\centering\arraybackslash}p{1.8cm}} 
	
		\toprule
        \centering \small \textbf{Rules}  
        & \parbox{1cm}{\centering \small \textbf{IDs}}
        & \parbox{1.8cm}{\centering \small \textbf{S-WC}}
        & \parbox{1.8cm}{\centering \small \textbf{PE-WC}}
        & \parbox{1.8cm}{\centering \small \textbf{M-WC}} \\
        
        \cmidrule(l){1-5}
        \footnotesize \multirow{6}{*}{16} &   \footnotesize 32  & \footnotesize 14.66\% & \footnotesize \multirow{6}{*}{14.23\%} & \footnotesize \multirow{6}{*}{15.85\%} \\
        
        \cmidrule(l){2-3}
         &   \footnotesize 64  & \footnotesize 15.85\% & & \\

        \cmidrule(l){2-3}
         &   \footnotesize 96  & \footnotesize 16.92\% & & \\

        \cmidrule(l){2-3}
         &   \footnotesize 128  & \footnotesize 18.07\% & & \\

         \cmidrule(l){1-5}
         
        \footnotesize \multirow{6}{*}{64} &   \footnotesize 32  & \footnotesize 15\% & \footnotesize \multirow{6}{*}{14.67\%} & \footnotesize \multirow{6}{*}{18.2\%} \\
        
        \cmidrule(l){2-3}
         &   \footnotesize 64  & \footnotesize 16.72\% & & \\

        \cmidrule(l){2-3}
         &   \footnotesize 96  & \footnotesize 18.3\% & & \\

        \cmidrule(l){2-3}
         &   \footnotesize 128  & \footnotesize 19.86\% & & \\

	\bottomrule
	\end{tabular}
	}
\end{table}

\section{Discussion}

The suitability of the evaluated primitives must be considered in light of key real-time requirements, including predictable access latency, scalability, hardware cost, and support for shared memory regions across initiators. In the following, we analyze these dimensions and highlight the trade-offs arising from the architectural differences between IOPMP, S-WC, and M-WC.

\mypara{Real-Time Suitability.} 
For real-time systems, suitability is primarily determined by access latency and predictability. Typical embedded workloads issue word-sized accesses to non-cached, memory-mapped peripherals, where even small sources of timing variability can significantly affect worst-case behavior. Both WC variants provide highly deterministic behavior: access checks have fixed latency due to parallel rule evaluation and permission checking, with no variable-depth structures or associative lookups. This enables straightforward worst-case analysis across configurations. In contrast, IOPMP exhibits configuration-dependent latency. While its best-case overhead is only three cycles, overall latency depends on the position of the matching rule within the configured MDs, complicating timing analysis and expanding the worst-case execution-time envelope, i.e., an important drawback for real-time and safety-critical systems. 

\mypara{Hardware Resources.} 
In terms of hardware cost, all three primitives exhibit similar scaling behavior with respect to the number of slots, which increases approximately linearly across designs (slightly super-linear for IOPMP). Among them, the S-WC consistently achieves the lowest resource usage. While absolute area figures depend on the target technology and should be interpreted cautiously, the relative scaling trends are more informative. The M-WC incurs a higher baseline cost due to the use of start–end registers by default. However, our preliminary results show that applying only the redesigned \textit{perm} field would reduce this overhead, aligning its footprint more closely with the S-WC. As the number of IIDs increases, the modified design becomes more resource-efficient beyond a crossover point, indicating that the proposed changes primarily benefit IID-rich configurations while preserving a compact footprint for smaller systems.

\mypara{Shareability.} 
Shareability differs significantly across the evaluated primitives due to their distinct rule-sharing models. The S-WC provides the greatest flexibility, allowing any number of IIDs to access a single rule. In contrast, the M-WC limits sharing to the number of available \textit{perm} fields per rule. In IOPMP, sharing depends on the MD configuration: an IID may access one or multiple MDs, and a rule is accessible only if it lies within the associated MD. Making a rule shareable across multiple IIDs may therefore require dedicating an MD to that rule or duplicating rules, increasing configuration complexity but incurring minimal area overhead. In our use cases, this limitation is mitigated by largely static configurations. Overall, S-WC offers the highest shareability, followed by IOPMP under favorable MD configurations, while M-WC provides the most constrained sharing. This limitation is partially mitigated through the General Read permission and accretive rule overlaying. Although M-WC lacks the fine-grained flexibility of the other designs, its reduced shareability is acceptable for typical consolidation scenarios (in particular for real-time automotive) and trades flexibility for improved scalability.

\mypara{Takeaways.}
This study yields three key observations. First, the WC variants are best suited for time-sensitive systems, as they provide low and fully deterministic access latency. Second, the proposed WC extensions improve scalability in deployments with large IID counts (approximately 46–128), reducing hardware overhead compared to the standard design. Third, these extensions trade some shareability for improved area efficiency, a balance that aligns well with the characteristics of horizontally integrated systems. Overall, these observations suggest that WC-based designs offer a strong foundation for real-time mixed-criticality systems.

\section{Related Work}

Prior work on system-level isolation has largely focused on mechanisms optimized for general-purpose platforms, most notably IOMMU-based designs that rely on page-table structures for permission enforcement~\cite{Malka2015, Markuze2016}. While effective in virtualized environments, these approaches introduce latency and timing variability that limit their suitability for real-time and embedded systems.

Within the RISC-V ecosystem, several efforts have explored variants of IOPMP-like mechanisms. An early prototype~\cite{Jien2022} presents an initial hardware–software implementation targeting an early draft of the IOPMP specification, providing a largely theoretical assessment of performance and area overhead. PROTEGO~\cite{protego} proposes a PMP-inspired protection unit that omits initiator identifiers and is designed to be placed between initiators and the interconnect, effectively extending PMP semantics to system initiators. A subsequent open-source implementation aligned with later drafts of the specification~\cite{Cunha2024} adopts sequential rule evaluation to minimize area, but incurs significantly higher access latency as a result. sIOPMP~\cite{siopmp} further extends the IOPMP model with mountable and remapping mechanisms to support large numbers of entries while sustaining system bandwidth; however, these extensions prioritize scalability and throughput over deterministic latency, limiting their applicability in real-time settings.

Although capability-based architectures (most notably CHERI~\cite{cheri}) were not explored in this work, they represent a complementary isolation paradigm. CHERI enforces fine-grained memory protection through unforgeable capabilities that encode bounds and permissions, enabling compartmentalization without page tables or target-side checkers. However, CHERI typically requires modifications to accelerators or substantial changes to the software stack, addressing a different layer of the isolation problem. Recent work has investigated importing CHERI capabilities into CHERI-unaware accelerators~\cite{cheri_capchecker}, but such approaches remain orthogonal to the protection mechanisms evaluated in this paper.

\section{Conclusion}

This paper evaluated target-side protection primitives for system-level isolation in heterogeneous SoCs, with a particular focus on real-time constraints. Through the implementation and comparison of IOPMP and WC variants in a CVA6-based system, we demonstrated how checker architecture directly influences access-latency predictability, scalability, and hardware cost. Overall, this work shows that World-based checkers provide deterministic and low-latency enforcement while scaling predictably with system size. The evaluated modifications can further reduce resource hardware consumption by up to 5\% in configurations with large identifier counts, without affecting latency. All artifacts will be open sourced to the community, and we expect these results to directly influence the evolution of RISC-V specifications and future RISC-V SoC designs.

\vfill

\end{document}